\documentclass[preprintnumbers,superscriptaddress,showkeys,byrevtex]{revtex4-2}
\usepackage{amsmath,amsfonts,amssymb,amscd,amsxtra,amsthm}
\usepackage{graphicx}
\usepackage{epstopdf}
\usepackage{bm}
\usepackage{orcidlink}
\usepackage{stackengine}
\usepackage{braket}
\usepackage{ragged2e}
\begin{document}
\preprint{PKNU-NuHaTh-2025}
\title{Entanglement entropy for $\pi^+ p$ elastic scattering using spin-density matrix}
%--------------------------------------------------
\author{Seung-il Nam$^*$\,\orcidlink{0000-0001-9603-9775}}
\email[Email: ]{sinam@pknu.ac.kr}
\affiliation{Department of Physics, Pukyong National University (PKNU), Busan 48513, Korea}
%--------------------------------------------------
\date{\today}
\begin{abstract}
We study the $\pi^+ p$ elastic scattering process using an effective Lagrangian approach that incorporates the $s$-, $u$-, and $t$-channel amplitudes, including $\Delta^{++}(1232)$, $\Delta^{0}(1232)$, neutron, and $\rho^0$ contributions. By constructing the spin-density matrices from the scattering amplitudes, we derive the von Neumann (entanglement) entropy associated with the spin degrees of freedom of the initial and final state particles. We compute the entropy by performing a partial trace over the spin subsystems, and its behavior is analyzed as a function of the center-of-mass energy $W$ and scattering angle $\theta$. We find that entropy exhibits nontrivial angular and energy dependences, reflecting the interplay among resonance contributions and background amplitudes. In particular, the $\Delta^{++}$ region shows relatively low entanglement, suggesting spin coherence dominated by the specific resonant contribution, while the background contributions increase entropy due to mixed spin configurations. These results indicate that entanglement entropy can serve as a novel probe into the quantum structure of hadronic scattering amplitudes, providing complementary insights beyond conventional cross-section analyses.
\end{abstract}
\keywords{von Neumann entropy, quantum entanglement, $\pi^+p$ elastic scattering, $\Delta$ resonance, effective Lagrangian approach, spin-density matrix.}
\maketitle
%--------------------------------------------------
\section{Introduction}
%--------------------------------------------------
Elastic scattering processes involving pions and nucleons provide essential insights into the non-perturbative regime of quantum chromodynamics (QCD) and the structure of baryonic resonances. Among them, the $\pi^+p$ elastic scattering is particularly notable due to the dominance of the $\Delta(1232)$ resonance at low energies, which plays a key role in understanding hadronic excitation mechanisms. Numerous experimental and theoretical studies have been dedicated to describing this process using an effective Lagrangian approach and partial wave analyses~\cite{ParticleDataGroup:2022pth, Pascalutsa:2006up}. While conventional analyses focus on cross-sections and resonance parameters, recent developments in quantum information theory have introduced new observables that probe deeper aspects of scattering dynamics. In particular, the von Neumann entropy, a measure of quantum entanglement, can be applied to hadronic reactions to study spin correlations and coherence~\cite{Amico:2007ag,Eisert:2008ur,Horodecki:2009zz}. This approach allows for an extended characterization of the scattering process beyond classical observables. Recent studies have directly linked entanglement entropy with particle creation in high-energy processes. Lin \textit{et al.}~\cite{Lin:2010pfa} investigated how expanding background fields generate entangled particle-antiparticle pairs, using von Neumann entropy to distinguish between entropy production and entanglement dynamics in quantum field theory. 

In this study, we investigate $\pi^+p \to \pi^+p$ elastic scattering using a tree-level effective Lagrangian approach that includes $s$-, $u$-, and $t$-channel contributions with the $\Delta^{++}$, $\Delta^0$, neutron, and $\rho^0$ intermediates. The primary reason for choosing this process is that all the involved particles can be traced experimentally because they carry an electric charge. We not only compute the total and differential cross-sections but also construct the spin-density matrix (SDM from canonical spin quantization. From the SDM, we extract the von Neumann entropy of the initial and final proton spin states to analyze quantum entanglement induced by hadronic interactions. We aim to examine how entropy behaves across different kinematic regimes and how it reflects the interplay between resonant and non-resonant contributions. By comparing the entropy distributions from various diagram combinations, we identify the quantum coherence structure encoded in the spin degrees of freedom. The results offer a novel quantum information-theoretic perspective on elastic scattering, paving the way for future studies of inelastic processes, polarization observables, and entangled baryonic states. This SDM formulation is not merely an imported formalism from quantum information theory, but rather a natural extension that arises from the intrinsic spin structure of the $\pi N$ scattering amplitude itself.

Specifically, the numerical results show that the total cross-section for $\pi^+p$ elastic scattering is strongly dominated by the $\Delta^{++}$ resonance near $W\approx 1230$~MeV, with background contributions from $\Delta^0$, neutron, and $\rho^0$ channels being significantly suppressed. The differential cross-section exhibits clear forward and backward peaking behavior at $W\approx 1285$~MeV, consistent with $p$-wave dominance. The SDM, constructed from the scattering amplitudes based on canonical spin quantization in the center-of-mass (c.m.) frame, reveals coherent spin structures in the resonance-dominated region and mixed configurations elsewhere. The von Neumann entropies $S_\mathrm{SDM}$ computed from reduced SDMs show strong angular and energy dependences, remaining relatively low in the region of $\Delta^{++}$ dominance and increasing when non-resonant amplitudes interfere. The entropy distributions for selected channel combinations demonstrate that background processes, such as neutron and $\rho^0$ exchange, substantially modify the spin-entanglement structure beyond the $\Delta^{++}$ region. In particular, they generate regions of both enhanced mixed-spin configurations and localized entropy suppression, depending on the interference pattern with the dominant $\Delta^{++}$ amplitude. These results provide quantum-informational insights into how resonant and non-resonant components redistribute spin coherence, going beyond what is visible in conventional cross-section observables. Beyond being a formal construction, the SDM and corresponding entanglement entropy are, in principle, experimentally reconstructible from polarization and spin-transfer observables in $\pi N$ scattering.

The remainder of this paper is organized as follows: In Sec.~II, we briefly describe the theoretical framework for the effective Lagrangian approach, which is used to compute the $\pi^+p$ elastic scattering process, and the construction of the von Neumann entropy from its SDMs. We present the numerical calculation results for the total and angular-dependent differential cross-sections, SDMs, and von Neumann entropies, along with detailed discussions, in Sec.~III.  Finally, Sec.~IV summarizes our conclusions and perspectives.
%--------------------------------------------------
\section{Theoretical Framework}
%--------------------------------------------------
%---------------------------------------------------------------------------------------------------
\subsection{Effective Lagrangians, scattering amplitudes, and form factors}
%---------------------------------------------------------------------------------------------------
\begin{figure}[t]
\includegraphics[width=16cm]{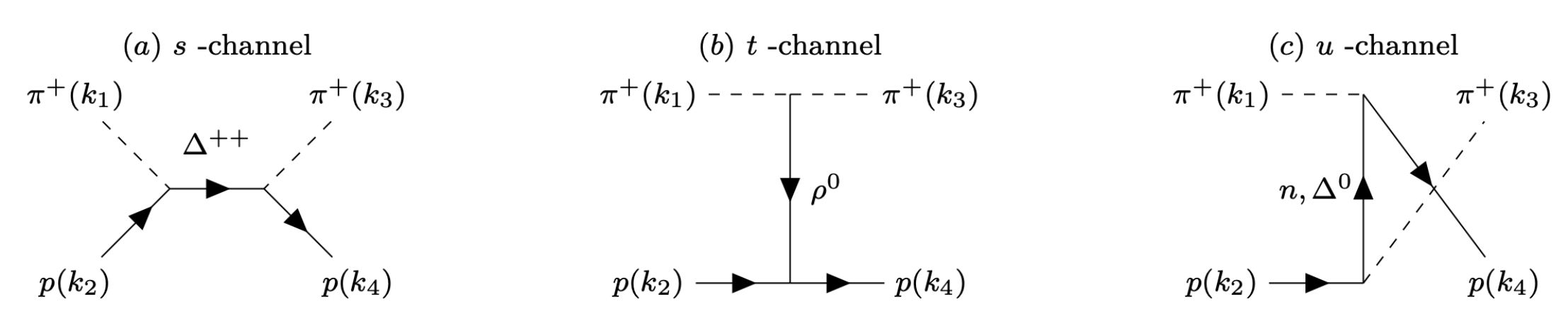}
\caption{Relevant Feynman diagrams for $\pi^+p$ elastic scattering: (a) $\Delta^{++}$-pole diagram in the $s$ channel, (b) $\rho^0$-exchange in the $t$ channel, and (c) $n$ and $\Delta^0$ exchanges in the $u$ channel. The four momenta for the involved particles are also defined by $k_{1,2,3,4}$.}
\label{FIG0}
\end{figure}
%--------------------------------------------------
In this section, we provide a brief introduction to the effective Lagrangian approach for the $\pi^+p$ elastic scattering process. Under the tree-level Born approximation, the relevant Feynman diagrams are depicted in Fig.~\ref{FIG0}: (a) $\Delta^{++}$-pole diagram in the $s$ channel, (b) $\rho^0$-exchange in the $t$ channel, and (c) $n$ and $\Delta^0$ exchanges in the $u$ channel. The solid and dashed lines represent the baryon (nucleon and $\Delta$) and pion, respectively. We also define the four momenta of the particles there. For the interaction vertices, we introduce the effective Lagrangians corresponding to each channel. The effective Lagrangian for the $\pi N \Delta$ vertex for Fig.~\ref{FIG0}(a) and (c) can be given in terms of the Rarita-Schwinger field formalism~\cite{Rarita:1941mf}:
%EQUATION
\begin{eqnarray}
\label{eq1}
\mathcal{L}_{\pi N \Delta}=\frac{f_{\pi N\Delta}}{M_\pi}\bar{\Delta}^{\mu}\partial_\mu (\bm{S}\cdot\bm{\pi})N + \mathrm{h.c},
\end{eqnarray}
%EQUATION
where $\Delta_\mu $ and $N$ are the $\Delta$ baryon and nucleon fields, respectively, and $\bm{S}$ the isospin transition operator between the isospin $1/2$ and $3/2$ fields. Thus, the scattering amplitudes are calculated straightforwardly as follows:
%EQUATION
\begin{eqnarray}
\label{eq2} 
i\mathcal{M}_s^{\Delta^{++}} = -\frac{if_{\pi N\Delta}^2}{M_\pi ^2}\bar{u}(k_4)k_{3\mu}\, G^{\mu\nu}(q_{1+2})\,k_{1\nu} u(k_2),\,\,\,
i\mathcal{M}_u^{\Delta^{0}}= -\frac{if_{\pi N\Delta}^2}{3M_\pi ^2}\bar{u}(k_4)k_{1\mu}\, 
G^{\mu\nu}(q_{2-3})\,k_{3\nu} u(k_2),
\end{eqnarray}
%EQUATION
where $q_{i\pm j}\equiv k_i\pm k_j$, and $f_{\pi N\Delta}$ denotes the coupling constant of $\pi N\Delta$. We employ the Mandelstam variables $s=q^2_{1+2}$ and $u=q^2_{2-3}$. $u(k_i)$ indicates the proton spinor in terms of the Bjorken-Drell (canonical) representation~\cite{BjorkenDrell1964}. The $\Delta$ baryon propagator $G^{\mu\nu}(q)$ in Eq.~(\ref{eq2}) is defined by
%EQUATION
\begin{eqnarray}
\label{eq3}
G^{\mu\nu}(q) = \frac{\rlap{/}{q}+M_\Delta}{q^2-M^2_\Delta + iM_\Delta\Gamma_\Delta}
\left[-g^{\mu\nu}+\frac{1}{3}\gamma^\mu\gamma^\nu+\frac{2q^\mu q^\nu}{3M^2_\Delta}-\frac{q^\mu\gamma^\nu-q^\nu\gamma^\mu}{3M_\Delta}\right].
\end{eqnarray}
%EQUATION

For Fig.~\ref{FIG0}(b), the relevant interaction Lagrangians are given by
%EQUATION>>>
\begin{eqnarray}
\mathcal{L}_{\rho\pi\pi} &=& g_{\rho\pi\pi} \vec{\rho}^\mu \cdot (\vec{\pi} \times \partial_\mu \vec{\pi}) 
= i g_{\rho\pi\pi} \rho^{0\mu} (\pi^+ \partial_\mu \pi^- - \pi^- \partial_\mu \pi^+), \\
\mathcal{L}_{\rho NN} &=& -g_{\rho NN} \bar{N} \left[\gamma^\mu \vec{\tau} \cdot \vec{\rho}_\mu 
- \frac{\kappa_\rho}{2M_N} \sigma^{\mu\nu} \vec{\tau} \cdot \partial_\nu \vec{\rho}_\mu \right] N.
\end{eqnarray}
%EQUATION>>>
Here, $\rho$ and $\pi$ denote the corresponding meson fields, and $\kappa_\rho$ denotes the anomalous magnetic moment of the $\rho$ meson. Then, we obtain the following scattering amplitude using the Mandelstam variable $t=q^2_{1-3}$:
%EQUATION>>>
\begin{eqnarray}
\mathcal{M}^{\rho^0}_t &=& -g_{\rho\pi\pi} g_{\rho NN}\bar{u}(k_4)
\left[\frac{q^\mu_{1+2} }{t- M_\rho^2} \right]
\left[ g_{\mu\nu} - \frac{q_{1-3\mu} q_{1-3\nu}}{M_\rho^2} \right]
\left[ \gamma^\nu +\frac{i\kappa_\rho}{2M_N} \sigma^{\nu\lambda} q_{1-3\lambda} \right] u(k_2).
\end{eqnarray}
%EQUATION>>>

Finally, for Fig.~\ref{FIG0}(c), the effective Lagrangian for the $\pi NN$ interaction in terms of the pseudo-vector scheme reads:
%EQUATION
\begin{eqnarray}
\label{eq4}
\mathcal{L}_{\pi NN} = -\frac{f_{\pi NN}}{M_\pi}\bar{N}\gamma_5 \rlap{/}{\partial}( \bm{\tau}\cdot\bm{\pi}) N+\mathrm{h.c.},
\end{eqnarray}
%EQUATION
where $f_{\pi NN}$ indicates the pseudovector $\pi NN$ coupling constant. Using the effective Lagrangian in Eq.~(\ref{eq4}), we calculate the scattering amplitude, resulting in
%EQUATION
\begin{eqnarray}
\label{eq5}
 i\mathcal{M}_u^n = -\frac{2if^2 _{\pi NN}}{3M_\pi ^2}\bar{u}(k_4)\gamma_5\rlap{/}{k}_1
\left[\frac{\rlap{/}{q}_{2-3}+M_N}{u -M_N^2}\right]\gamma_5\rlap{/}{k}_3u(k_2).
\end{eqnarray}
%EQUATION
The spin-averaged differential cross-section, summing all the ($s,t,u$)-channel scattering amplitudes, reads
%EQUATION
\begin{eqnarray}
\label{eq8}
\frac{d\sigma_{\pi^+p}}{d\Omega} &=& \frac{1}{128\pi^{2}s}\frac{|\textbf{k}_3|}{|\textbf{k}_1|}
\sum_\mathrm{spin}\left|\mathcal{M}^\mathrm{total}_{\pi^+p}\right|^2,
\end{eqnarray}
%EQUATION
where $\textbf{k}_1$ and $\textbf{k}_3$ represent the c.m. three momenta of the particles in the initial and final states, respectively, and the total amplitude with the phenomenological form factors can be written as
%EQUATION
\begin{eqnarray}
\label{eq9}
i\mathcal{M}^\mathrm{total}_{\pi^+p}= i \mathcal{M}_s^{\Delta^{++}}F_s^{\Delta^{++}}
+i \mathcal{M}^{\rho^0}_t F_{\mathrm{t}}+\left[i \mathcal{M}_u^{\Delta^{0}}F_u^{\Delta^{0}}
+i \mathcal{M}_u^nF_u^n\right].
\end{eqnarray}
%EQUATION
The form factors, which are responsible for incorporating the spatial extension of the hadrons, are defined for each channel by
%EQUATION
\begin{eqnarray}
\label{eq10}
F^h_{x}(x) &=& \frac{\Lambda^{4}_\mathrm{cutoff}}{\Lambda^{4}_\mathrm{cutoff}+(x-M^{2}_h)^{2}},
\end{eqnarray}
%EQUATION
where $x$ and $h$ denote the Mandelstam variable ($s,t,u$) and corresponding off-shell hadrons, respectively. All the relevant parameters, including couplings and cutoff $\Lambda$, will be discussed in Sec.~III. 

%-------------------------------------------------
\section{Spin-density matrix and entanglement entropy formalism}
%-------------------------------------------------
In this work, quantum spin entanglement in $\pi^+p$ elastic scattering is analyzed through the SDM formalism. The dynamical information contained in the scattering amplitude is mapped to a density operator acting on the spin degrees of freedom, enabling a quantum-information interpretation of hadronic spin correlations.
%-------------------------------------------------
\subsection{Spin basis and SDM construction}
%-------------------------------------------------
The scattering amplitudes are written as
\begin{equation}
M_{s_i s_f} \equiv \langle s_f | \mathcal{M} | s_i \rangle,
\qquad s_{i,f} = \uparrow,\downarrow,
\end{equation}
where $s_i$ ($s_f$) labels the spin of the initial (final) proton. We adopt the canonical spin basis, in which spin is quantized in the rest frame along a fixed direction, and the spinor is subsequently boosted to the c.m. frame. This is in contrast to the helicity basis, where spin is quantized along the momentum direction; the two bases differ by a momentum-dependent Wigner rotation. Consequently, parity constraints take different explicit forms in the two bases.

For the amplitude matrix
\begin{equation}
\mathcal{M}_{s_i s_f}
= (\mathcal{M}_{\uparrow\uparrow},
\mathcal{M}_{\uparrow\downarrow},
\mathcal{M}_{\downarrow\uparrow},
\mathcal{M}_{\downarrow\downarrow})
\equiv (a,b,c,d),
\end{equation}
the $4\times 4$ SDM for the proton spin system is constructed as
\begin{equation}
\rho_{\mathrm{SDM}}
= \frac{1}{\mathcal{N}}
\begin{pmatrix}
|a|^2 & ab^* & ac^* & ad^* \\
a^*b  & |b|^2 & bc^* & bd^* \\
a^*c  & b^*c  & |c|^2 & cd^* \\
a^*d  & b^*d  & c^*d  & |d|^2
\end{pmatrix},
\qquad
\mathcal{N} = |a|^2+|b|^2+|c|^2+|d|^2.
\label{eq:rhoSDM}
\end{equation}
By construction, $\mathrm{Tr}[\rho_{\mathrm{SDM}}]=1$. The diagonal elements encode 
spin-state probabilities, while the off-diagonal elements capture quantum 
coherence between spin configurations.

In the helicity basis and in the c.m.~frame, parity relates the amplitudes via $a=d$ and $b=-c$. In the canonical basis, however, these equalities do not hold element-wise due to Wigner rotations. Nevertheless, parity still guarantees that 
only two independent physical amplitudes remain, ensuring consistency with the symmetry structure of $\pi N$ scattering. The canonical basis preserves rest-frame spin alignment under boosts, producing Wigner rotations. Therefore, parity relations differ from the helicity basis at the matrix-element level, while the number of independent amplitudes remains two by symmetry. For more detailed explanations, see the Appendix.

%-------------------------------------------------
\subsection{Reduced density matrices and von Neumann entropy}
%-------------------------------------------------
To quantify spin entanglement, we trace over either the initial or final spin
index to obtain a $2\times2$ reduced SDM:
\begin{equation}
\rho_{\mathrm{SDM}}^{i}
= \mathrm{Tr}_f (\rho_{\mathrm{SDM}})
= \frac{1}{\mathcal{N}}
\begin{pmatrix}
|a|^2+|b|^2 & ac^*+bd^* \\
a^*c+b^*d  & |c|^2+|d|^2
\end{pmatrix},
\qquad
\rho_{\mathrm{SDM}}^{f}
= \mathrm{Tr}_i (\rho_{\mathrm{SDM}})
= \frac{1}{\mathcal{N}}
\begin{pmatrix}
|a|^2+|c|^2 & ab^*+cd^* \\
a^*b+c^*d  & |b|^2+|d|^2
\end{pmatrix}.
\label{eq:rho_reduced}
\end{equation}

The eigenvalues of $\rho_{\mathrm{SDM}}^{i,f}$ are
\begin{equation}
\lambda_{\pm}^{i,f}
= \frac{1}{2}\left(1 \pm \sqrt{1-4\,\mathrm{Det}[\rho_{\mathrm{SDM}}^{i,f}]}\right),
\label{eq:eigenvalues}
\end{equation}
leading to the von Neumann entropy
\begin{equation}
S_{\mathrm{SDM}}^{i,f}
= -\sum_{\alpha=\pm}
\lambda_{\alpha}^{i,f}
\log \lambda_{\alpha}^{i,f}.
\end{equation}

Here, $S_{\mathrm{SDM}}$ quantifies the degree of entanglement between proton spin and all other degrees of freedom (orbital motion, isospin, etc.). A vanishing entropy indicates a pure spin state with no entanglement, whereas a nonzero
entropy reflects mixing induced by the scattering dynamics. A comparison of $S_{\mathrm{SDM}}^{i}$ and $S_{\mathrm{SDM}}^{f}$ reveals how quantum information is redistributed between initial and final spin subsystems during the elastic process. We note that the SDM elements $\rho_{ij}$ are not abstract quantities: in elastic $\pi N$ scattering, they correspond to measurable polarization and spin-transfer observables, such as the single-spin asymmetry $P$ ($A_y$), and double-polarization coefficients $D_{ij}$ and $C_{ij}$. Thus, $\rho_{ij} \leftrightarrow (P, A_y, D_{ij}, C_{ij})$ provides an operational bridge to experimental analysis, ensuring that the entropy constructed from $\rho$ remains, in principle, experimentally reconstructible~\cite{Fano:1957zz,Bystricky1978,Arndt1986,Arndt2006,Meyer2014}.

%-------------------------------------------------
\section{Numerical results and Discussions}
%-------------------------------------------------
%-------------------------------------------------
\subsection{Cross-Sections and Angular Distributions}
%-------------------------------------------------
%FIGURE
\begin{figure}[ht]
\centering
\topinset{(a)}{\includegraphics[width=7.5cm]{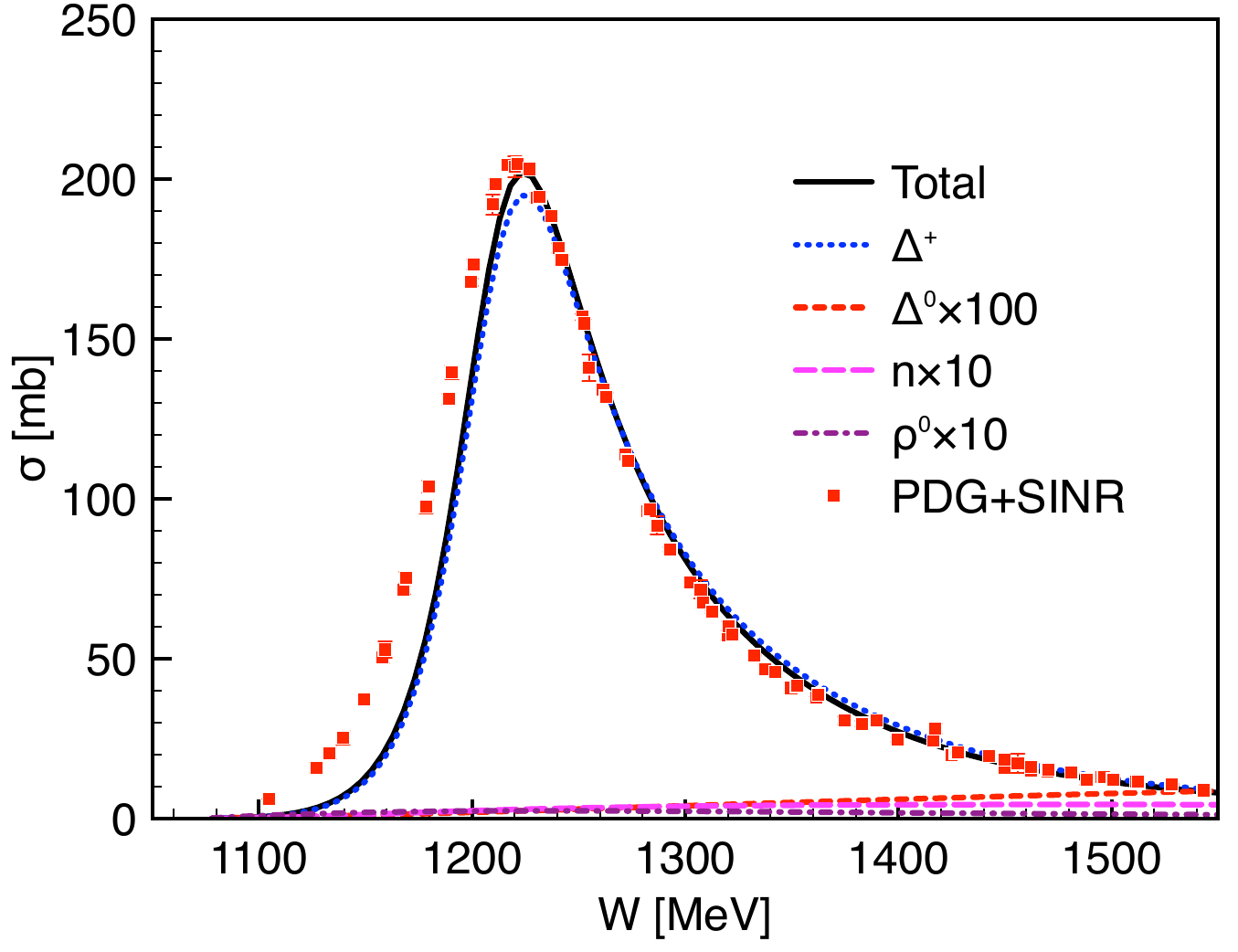}}{-0.3cm}{0.5cm}
\hspace{0.5cm}
\topinset{(b)}{\includegraphics[width=7.5cm]{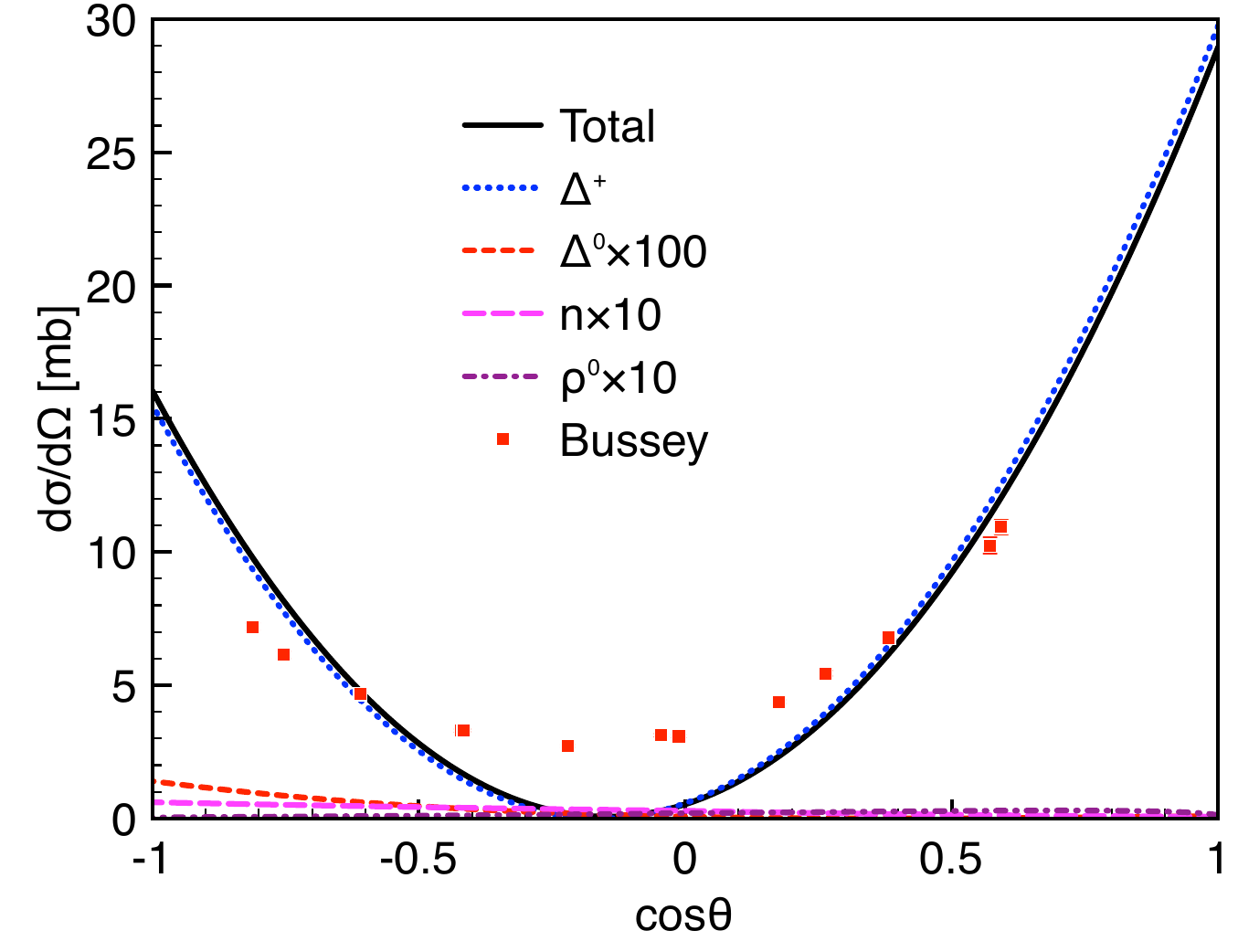}}{-0.3cm}{0.5cm}
\topinset{(c)}{\includegraphics[width=9.5cm]{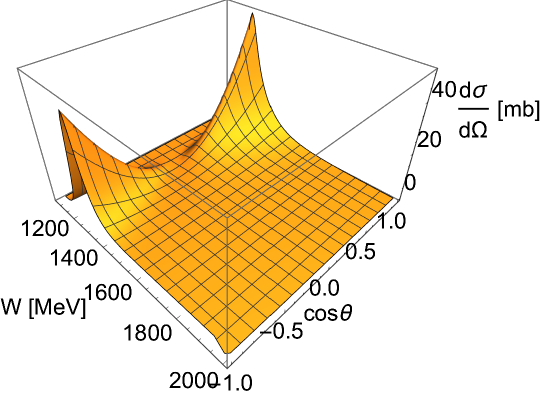}}{-0.3cm}{0.5cm}
\caption{\justifying (Color online) (a) Total cross-section for the $\pi^+p$ elastic scattering process as a function of c.m. energy $W$ [MeV]. We also show separate contributions and experimental data from PDG and SINR~\cite{PDGplot,Pedroni:1978it}. (b) Angular-dependent differential cross-section as a function of the scattering angle $\theta$ of the outgoing $\pi^+p$ in the c.m. frame at $W\approx1.285$ GeV with the experimental data from Ref.~\cite{Bussey:1973gz}. (c) Differential cross-section as a function of $W$ and $\cos\theta$.}
\label{FIG12}
\end{figure}
%FIGURE
In this section, we present the numerical results and their corresponding discussions. In Table~\ref{TAB1}, we list all the relevant parameters for the numerical calculations from experimental~\cite{ParticleDataGroup:2024cfk} and theoretical information~\cite{Han:2024hzx}. 

%TABLE
\begin{table}[b]
\begin{tabular}{c|c|c|c|c||c|c|c|c|c} 
$f_{\pi NN}$&$f_{\pi N\Delta}$&$g_{\rho\pi\pi}$&$g_{\rho NN}$
&$\kappa_\rho$&$M_p$&$(M,\Gamma)_{\Delta^{++,0}}$&$M_{\pi^+}$
&$(M,\Gamma)_{\rho^0}$&$\Lambda_\mathrm{cutoff}$\\
\hline
$0.989$&$2.2$&$6.0$&$3.25$
&$6.1$&$939.272$ MeV&$(1210,80)$ MeV&$139.570$ MeV&$(763,72.5)$ MeV&$750$ MeV
\end{tabular}
\caption{\justifying Relevant parameters for the numerical calculations.}
\label{TAB1}
\end{table}
%TABLE
Note that some masses and widths of the hadrons are modified to fit the data, e.g., those for $\Delta$. More detailed computations and discussions on the model can be found in our previous work~\cite{Han:2021zrh}.

Fig.~\ref{FIG12}(a) shows the total cross-section $\sigma$ as a function of the c.m. energy $W$ [MeV]. The $\Delta^{++}$ resonance dominates in the region around $W\approx 1230$~MeV, producing a sharp peak that aligns well with experimental data from PDG and SINR~\cite{PDGplot,Pedroni:1978it}. The background contributions from $\Delta^0$, neutron, and $\rho^0$ exchanges are significantly suppressed but are included to provide completeness and to study their effects on entanglement properties. The theoretical predictions qualitatively reproduce the energy dependence of the cross-section. This result confirms that the $\Delta$ resonance captures the essential dynamics of low-energy $\pi^+p$ scattering.

To investigate the angular dependence of the reaction, Fig.~\ref{FIG12}(b) presents the differential cross-section $d\sigma/d\Omega$ at $W \approx 1285$~MeV as a function of the scattering angle $\theta$ in the c.m. frame. The result exhibits a pronounced forward-backward asymmetry, consistent with dominant $p$-wave contributions from the $\Delta$ resonance in the $s$ channel. The agreement with experimental data, particularly in the forward region, is encouraging and validates the underlying effective interaction vertices. The angular structure arises from interference among multiple spin-parity amplitudes and plays a key role in shaping the spin-entanglement structure discussed below.

In Fig.~\ref{FIG12}(c), we provide a two-dimensional visualization of the differential cross-section as a function of $W$ and $\cos\theta$, offering a comprehensive picture of the interplay between energy and angle across the scattering phase space. The plot reveals regions of enhanced cross-section corresponding to the $\Delta$ resonance and highlights the forward and backward peaking behaviors. Outside the resonance region, the cross-section becomes relatively smooth and flat. This information is crucial in identifying kinematic domains where spin coherence or decoherence may occur due to underlying dynamical mechanisms.

%-------------------------------------------------
\subsection{Spin-Density Matrices and Entanglement Entropy}
%-------------------------------------------------
%FIGURE
\begin{figure}[t]
\begin{tabular}{cccc}
\topinset{Re$[\rho_{11}]$}{\includegraphics[width=2.0cm]{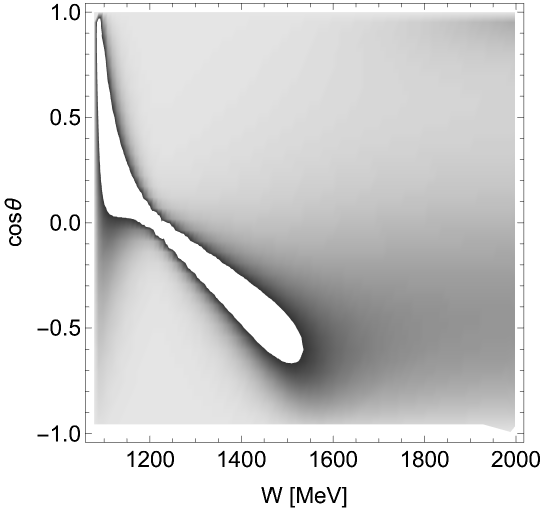}}{-0.3cm}{0.2cm}
\topinset{Re$[\rho_{12}]$}{\includegraphics[width=2.0cm]{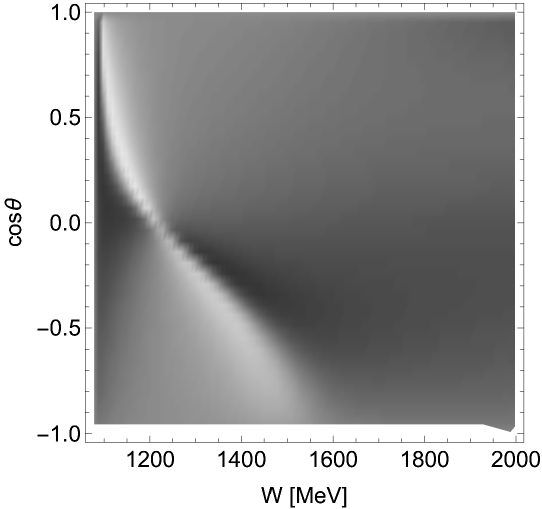}}{-0.3cm}{0.2cm}
\topinset{Re$[\rho_{13}]$}{\includegraphics[width=2.0cm]{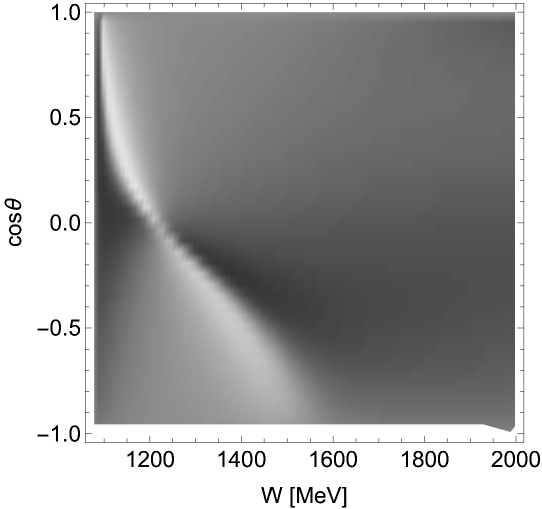}}{-0.3cm}{0.2cm}
\topinset{Re$[\rho_{14}]$}{\includegraphics[width=2.0cm]{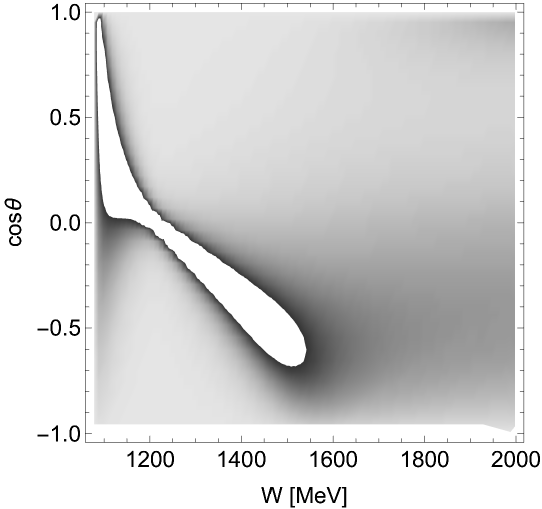}}{-0.3cm}{0.2cm}
\hspace{0.2cm}
\topinset{Im$[\rho_{11}]$}{\includegraphics[width=2.0cm]{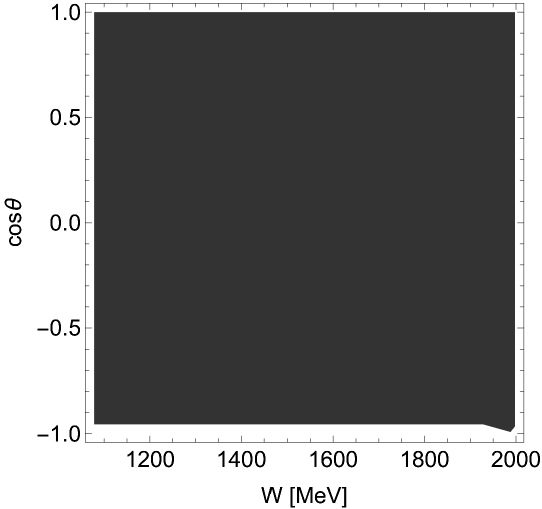}}{-0.3cm}{0.2cm}
\topinset{Im$[\rho_{12}]$}{\includegraphics[width=2.0cm]{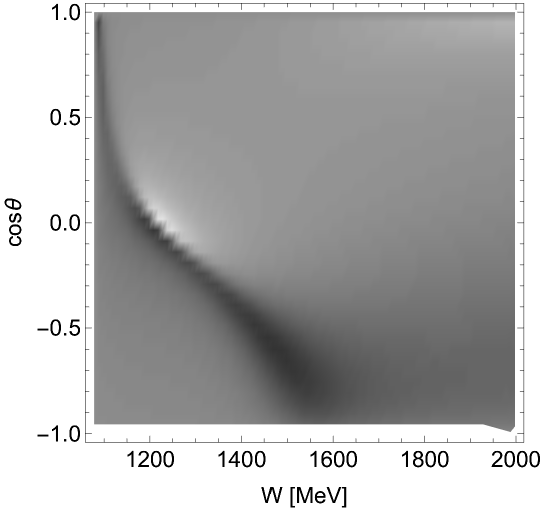}}{-0.3cm}{0.2cm}
\topinset{Im$[\rho_{13}]$}{\includegraphics[width=2.0cm]{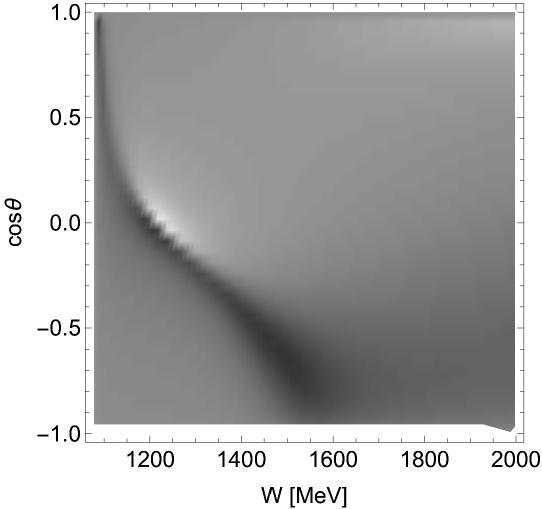}}{-0.3cm}{0.2cm}
\topinset{Im$[\rho_{14}]$}{\includegraphics[width=2.0cm]{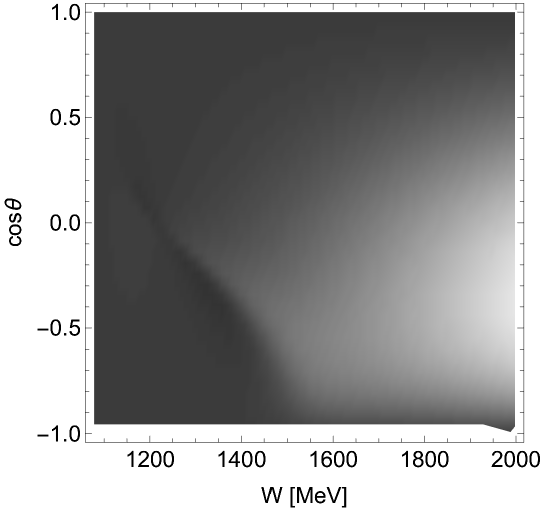}}{-0.3cm}{0.2cm}
\end{tabular}
\begin{tabular}{cccc}
\topinset{Re$[\rho_{21}]$}{\includegraphics[width=2.0cm]{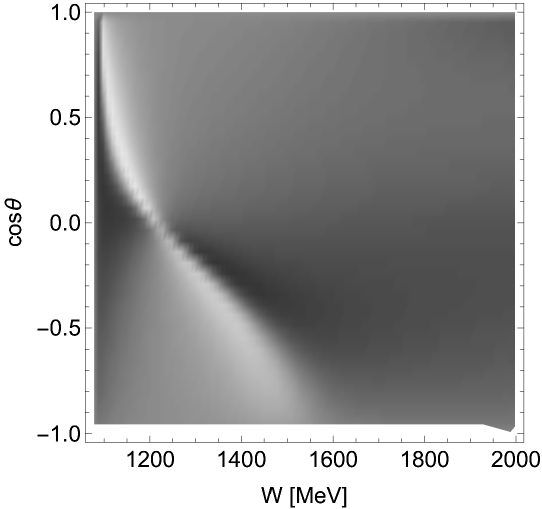}}{-0.3cm}{0.2cm}
\topinset{Re$[\rho_{22}]$}{\includegraphics[width=2.0cm]{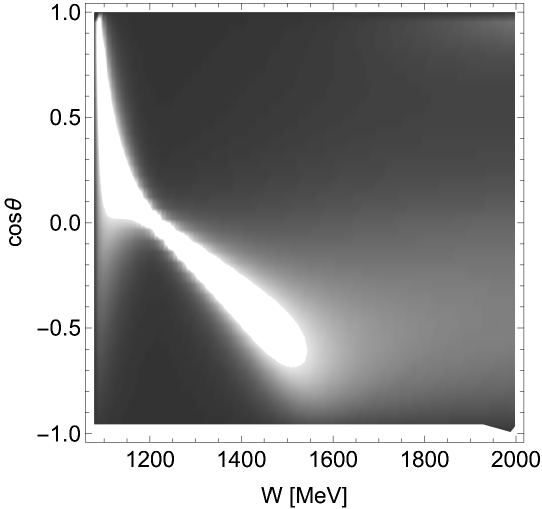}}{-0.3cm}{0.2cm}
\topinset{Re$[\rho_{23}]$}{\includegraphics[width=2.0cm]{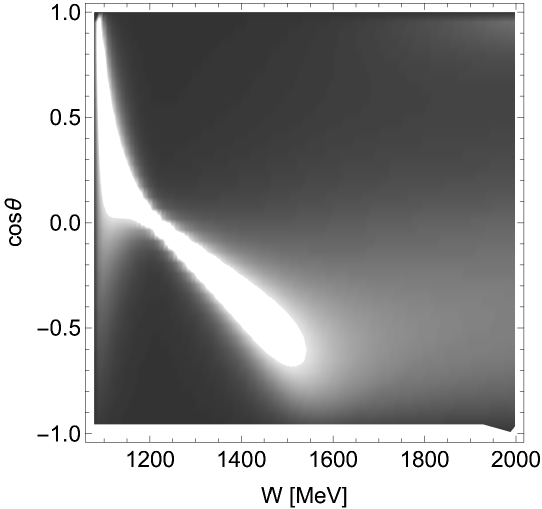}}{-0.3cm}{0.2cm}
\topinset{Re$[\rho_{24}]$}{\includegraphics[width=2.0cm]{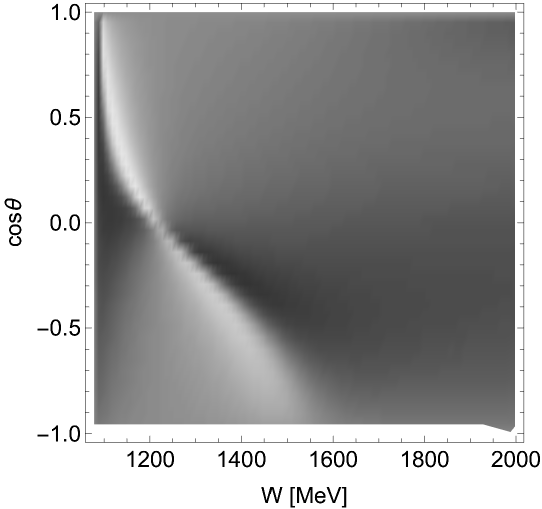}}{-0.3cm}{0.2cm}
\hspace{0.2cm}
\topinset{Im$[\rho_{21}]$}{\includegraphics[width=2.0cm]{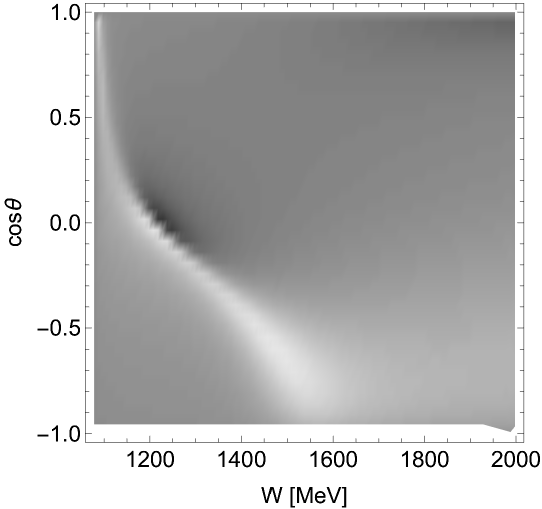}}{-0.3cm}{0.2cm}
\topinset{Im$[\rho_{22}]$}{\includegraphics[width=2.0cm]{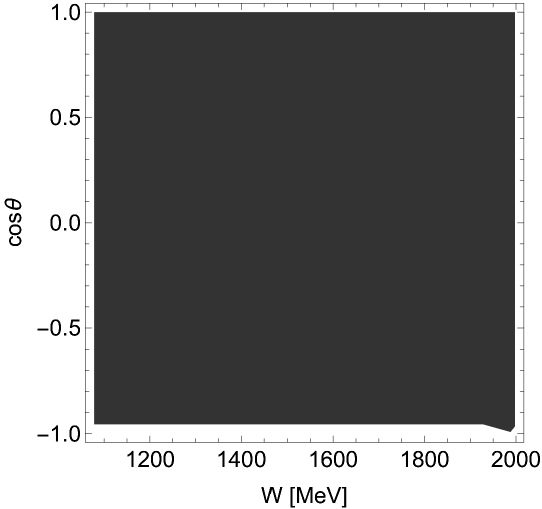}}{-0.3cm}{0.2cm}
\topinset{Im$[\rho_{23}]$}{\includegraphics[width=2.0cm]{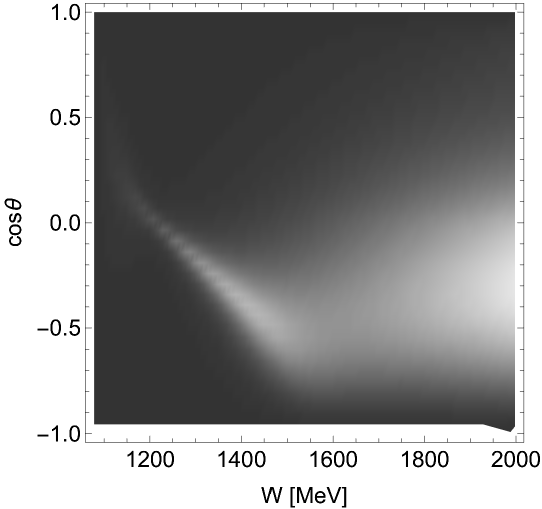}}{-0.3cm}{0.2cm}
\topinset{Im$[\rho_{24}]$}{\includegraphics[width=2.0cm]{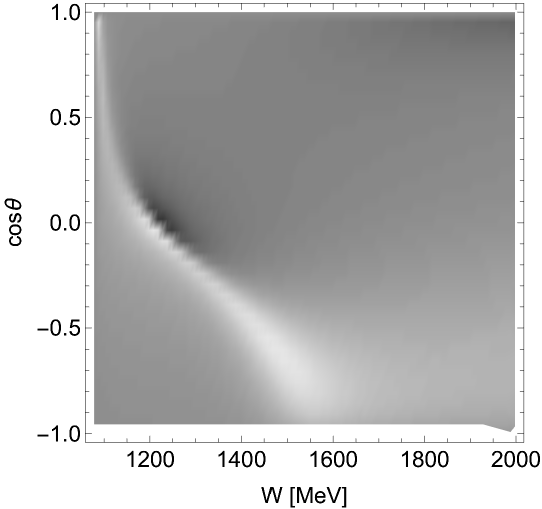}}{-0.3cm}{0.2cm}
\end{tabular}
\begin{tabular}{cccc}
\topinset{Re$[\rho_{31}]$}{\includegraphics[width=2.0cm]{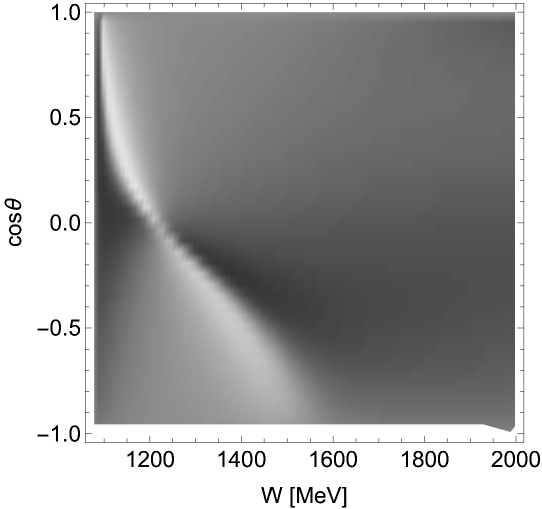}}{-0.3cm}{0.2cm}
\topinset{Re$[\rho_{32}]$}{\includegraphics[width=2.0cm]{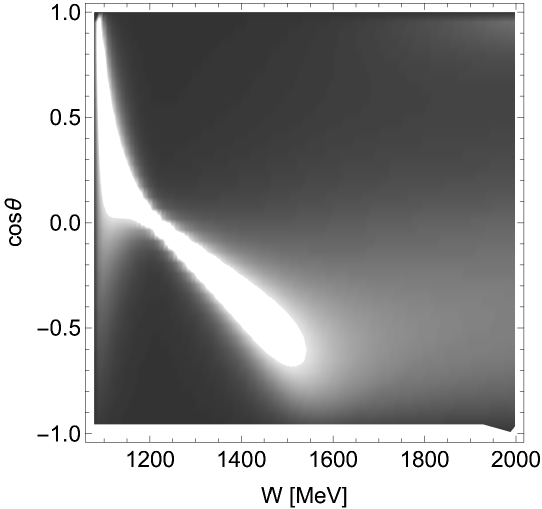}}{-0.3cm}{0.2cm}
\topinset{Re$[\rho_{33}]$}{\includegraphics[width=2.0cm]{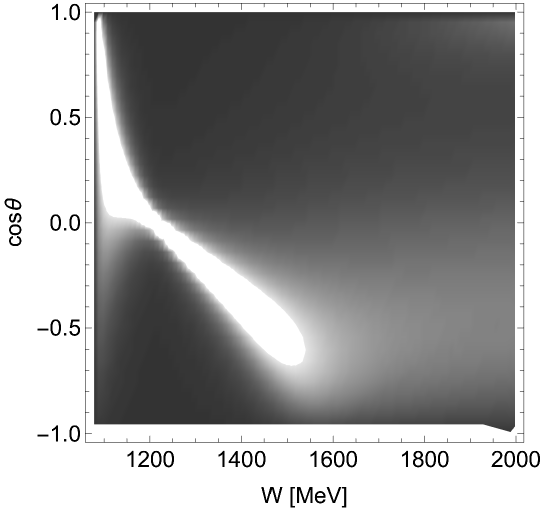}}{-0.3cm}{0.2cm}
\topinset{Re$[\rho_{34}]$}{\includegraphics[width=2.0cm]{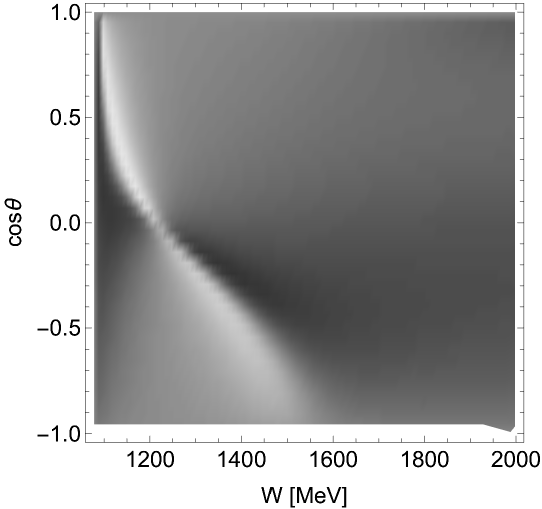}}{-0.3cm}{0.2cm}
\hspace{0.2cm}
\topinset{Im$[\rho_{31}]$}{\includegraphics[width=2.0cm]{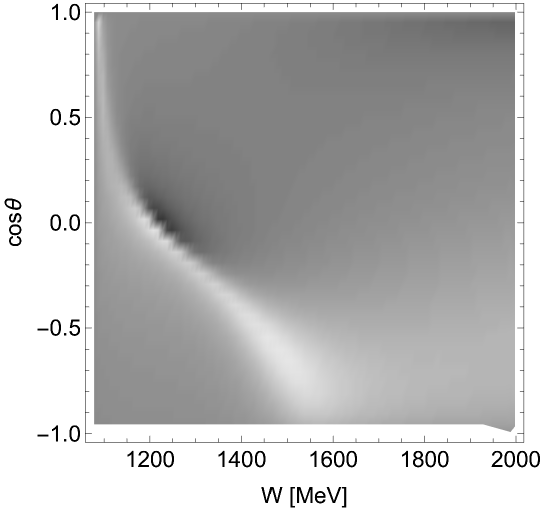}}{-0.3cm}{0.2cm}
\topinset{Im$[\rho_{32}]$}{\includegraphics[width=2.0cm]{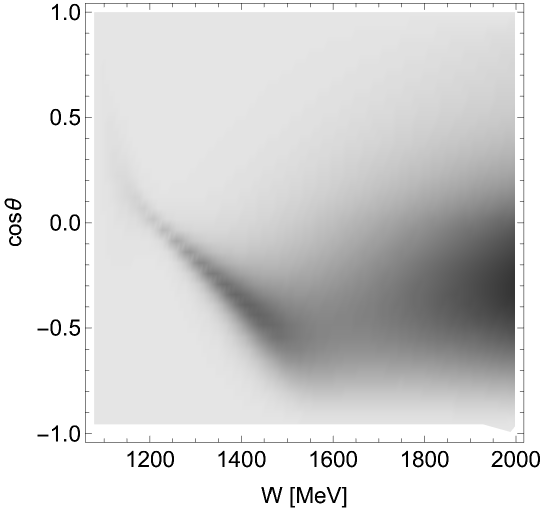}}{-0.3cm}{0.2cm}
\topinset{Im$[\rho_{33}]$}{\includegraphics[width=2.0cm]{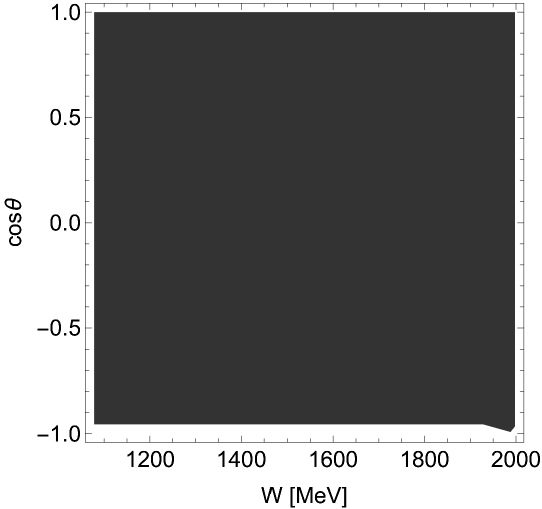}}{-0.3cm}{0.2cm}
\topinset{Im$[\rho_{34}]$}{\includegraphics[width=2.0cm]{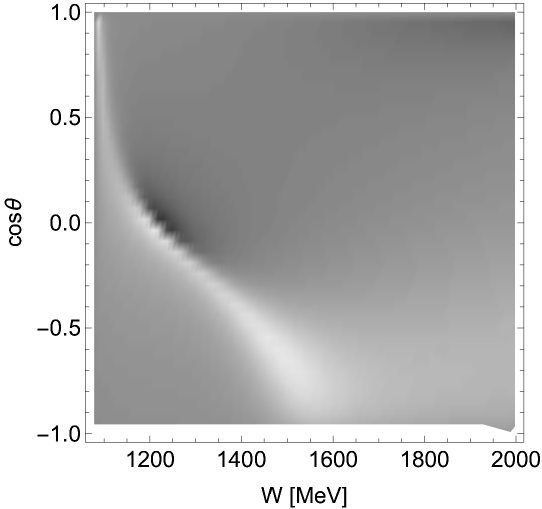}}{-0.3cm}{0.2cm}
\end{tabular}
\begin{tabular}{cccc}
\topinset{Re$[\rho_{41}]$}{\includegraphics[width=2.0cm]{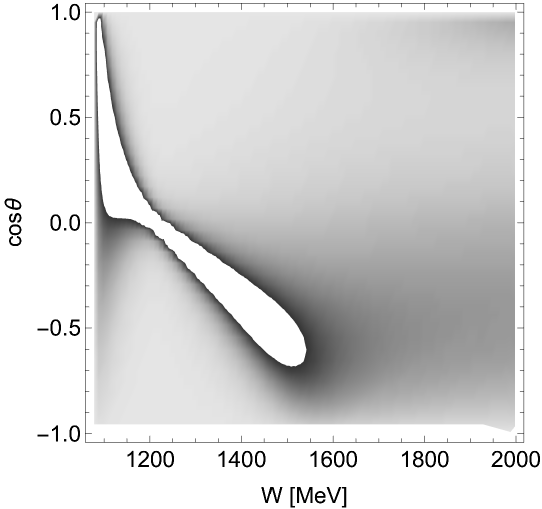}}{-0.3cm}{0.2cm}
\topinset{Re$[\rho_{42}]$}{\includegraphics[width=2.0cm]{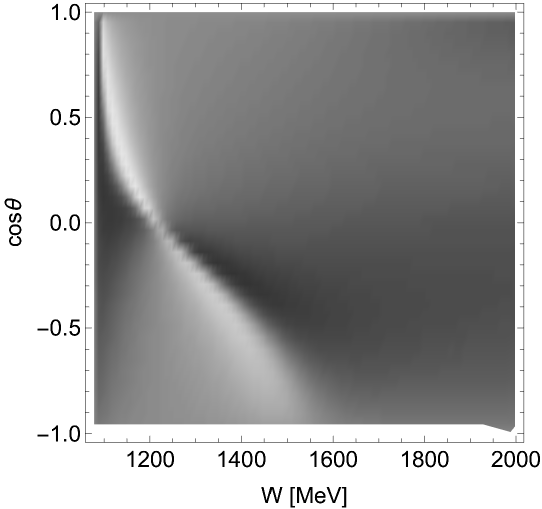}}{-0.3cm}{0.2cm}
\topinset{Re$[\rho_{43}]$}{\includegraphics[width=2.0cm]{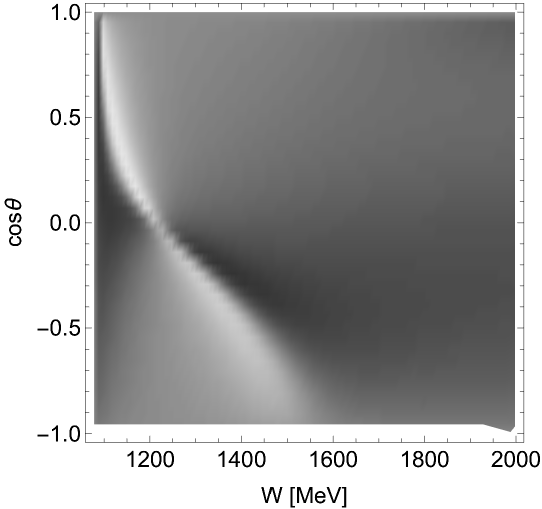}}{-0.3cm}{0.2cm}
\topinset{Re$[\rho_{44}]$}{\includegraphics[width=2.0cm]{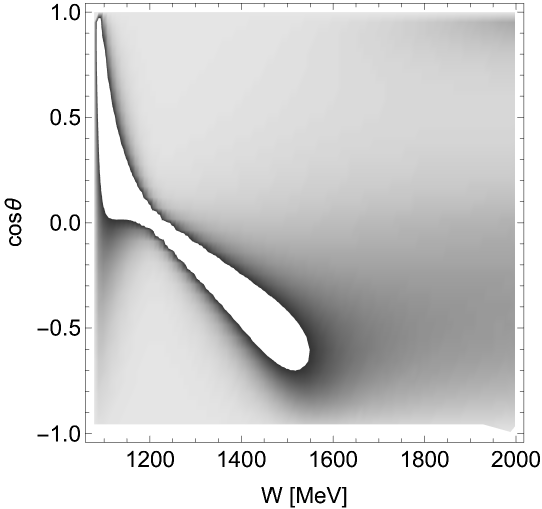}}{-0.3cm}{0.2cm}
\hspace{0.2cm}
\topinset{Im$[\rho_{41}]$}{\includegraphics[width=2.0cm]{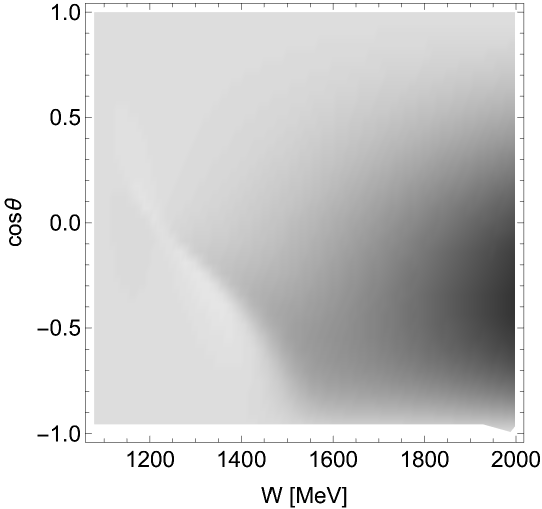}}{-0.3cm}{0.2cm}
\topinset{Im$[\rho_{42}]$}{\includegraphics[width=2.0cm]{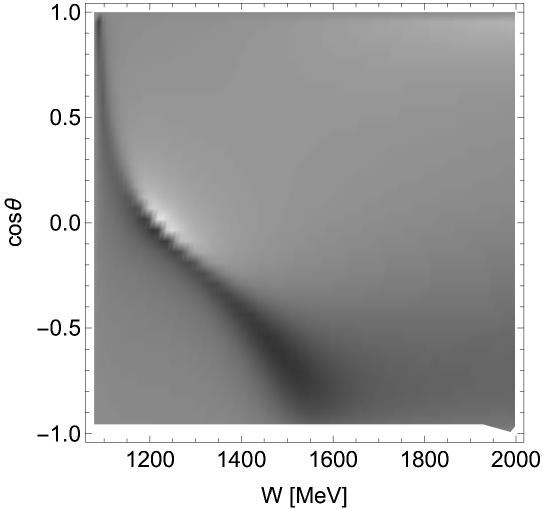}}{-0.3cm}{0.2cm}
\topinset{Im$[\rho_{43}]$}{\includegraphics[width=2.0cm]{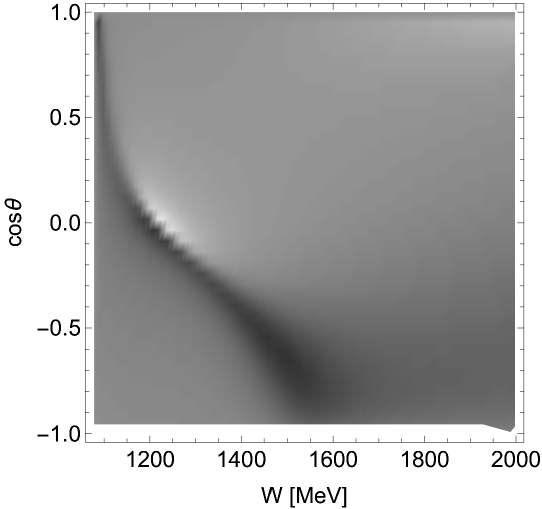}}{-0.3cm}{0.2cm}
\topinset{Im$[\rho_{44}]$}{\includegraphics[width=2.0cm]{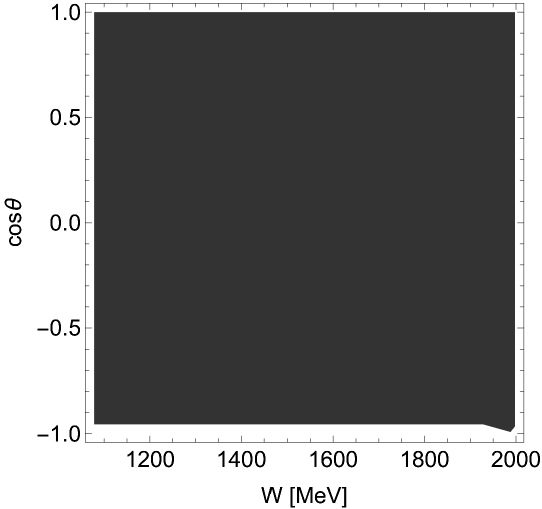}}{-0.3cm}{0.2cm}
\end{tabular}
\caption{\justifying  Real and imaginary matrix elements of $\rho_\mathrm{SDM}$ in separate panels as functions of $W=(W_\mathrm{th}-2.0)$ GeV and $|\cos\theta|\leq1$ for the horizontal and vertical axes, respectively. $W_\mathrm{th}=M_{\pi^+}+M_p$ denotes the c.m. threshold energy.}
\label{FIG3}
\end{figure}
%FIGURE
To explore the quantum spin structure, we construct the $4 \times 4$ SDM $\rho_{\text{SDM}}$ for the initial and final state protons. Fig.~\ref{FIG3} presents the real and imaginary parts of all matrix elements as functions of $W$ (horizontal axis) and $\cos\theta$ (vertical axis). The matrix is normalized so that its trace is unity and encapsulates the probability distributions and coherences of spin transitions during scattering. The physical meaning of these elements is crucial: the diagonal elements $\text{Re}[\rho_{ii}]$ quantify the probability of finding the proton in a specific spin state, while the off-diagonal elements ($\text{Re}[\rho_{ij}]$ and $\text{Im}[\rho_{ij}]$ for $i \neq j$) signify the presence and strength of quantum interference between different spin states, indicating the degree of coherence or entanglement. Note that the diagonal elements of $\mathrm{Im}[\rho_{\text{SDM}}]$ are zero by construction, as probabilities are real. Result interpretation shows that variations in these elements across kinematic regions signal transitions between coherent and mixed spin configurations; for instance, a region with dominant diagonal elements and small off-diagonal elements suggests high spin coherence, often associated with a single dominant scattering channel, whereas a significant presence of non zero off-diagonal elements implies a mix of spin states and active interference, reflecting the presence of a background, as in the present work, or the interplay of multiple scattering amplitudes.

%FIGURE
\begin{figure}[t]
\topinset{(a) $\Delta^{++}+\Delta^0+n+\rho^0$}{\includegraphics[width=7.5cm]{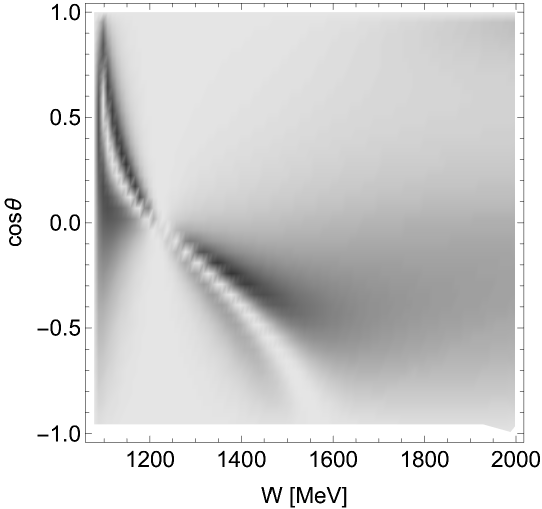}}{-0.4cm}{0.2cm}
\hspace{0.5cm}
\topinset{(b) $\Delta^{++}+\Delta^0+n+\rho^0$}{\includegraphics[width=7.5cm]{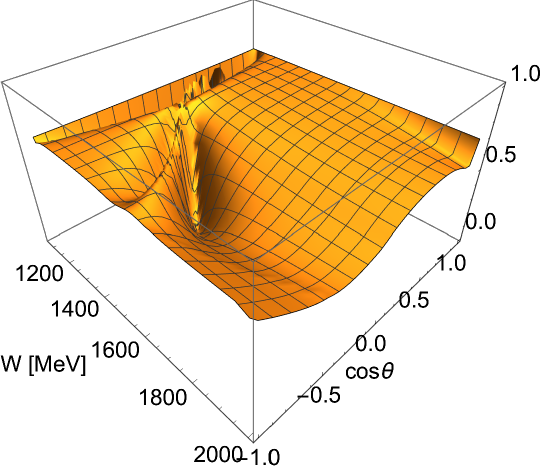}}{-0.4cm}{0.2cm}
\caption{\justifying (Color online) von Neumann entropy for the SDM, i.e., $S_\mathrm{SDM}$ as a function of $W$ [MeV] and $\cos\theta$, including all the contributions, $\Delta^{++}$, $\Delta^0$, $n$, and $\rho^0$.}
\label{FIG4}
\end{figure}
%FIGURE
The von Neumann entropies $S^{i,f}_{\text{SDM}}$ for the initial and final spin subsystems are calculated from the eigenvalues of their respective reduced density matrices, obtained via partial trace operations. We find that the von Neumann entropies of the initial and final spin subsystems are numerically identical across the entire kinematic domain: {\color{blue}$S^i_{\text{SDM}}=S^f_{\text{SDM}}=S_{\text{SDM}}$}. This implies that the degree of spin entanglement with other degrees of freedom, such as momentum or isospin, is conserved during the elastic scattering process. Physically, this indicates that the scattering acts as a unitary transformation on the spin space, redistributing quantum information but neither generating nor destroying it. For a pure bipartite state, the reduced density matrices of the two subsystems have identical nonzero eigenvalues~\cite{NielsenChuang2000}. Thus, for elastic scattering with a pure initial state, we have $S_{\mathrm{SDM}}^i = S_{\mathrm{SDM}}^f$, consistent with unitary evolution in the spin Hilbert space.

Fig.~\ref{FIG4} displays the entropy value as a function of $W$ and $\cos\theta$, including contributions from all relevant diagrams. The entropy presents a nontrivial structure in the $\Delta^{++}$-dominant region, indicating that complicated (dis)entanglement occurs from interference with different spin contributions ($n$ and $\rho^0$). For instance, a small but finite forward scattering enhancement is generated by the $\rho^0$ contribution. In contrast, beyond the $\Delta^{++}$ region, the entropy increases and becomes relatively flat, where multiple amplitudes overlap and interfere strongly, i.e., stronger entanglements, especially those involving background channels such as $n$ and $\rho^0$. This clearly shows how spin entanglement responds to the dynamical structure of the scattering amplitude. As discussed in detail below, the nontrivial dip (valley-like) suggests that particles with different spins are mixed by lowering the entropy, i.e., \textit{disentanglement}. 

%FIGURE
\begin{figure}[t]
\begin{tabular}{ccc}
\topinset{(a) $\Delta^{++}+\Delta^0$}{\includegraphics[width=4.5cm]{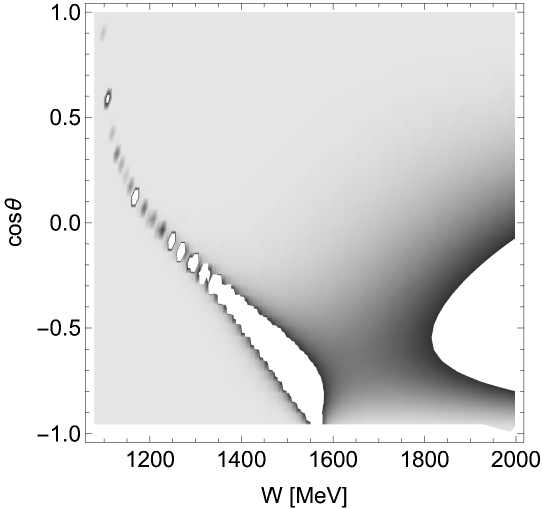}}{-0.4cm}{0.2cm}
\hspace{0.5cm}
\topinset{(b) $\Delta^{++}+n$}{\includegraphics[width=4.5cm]{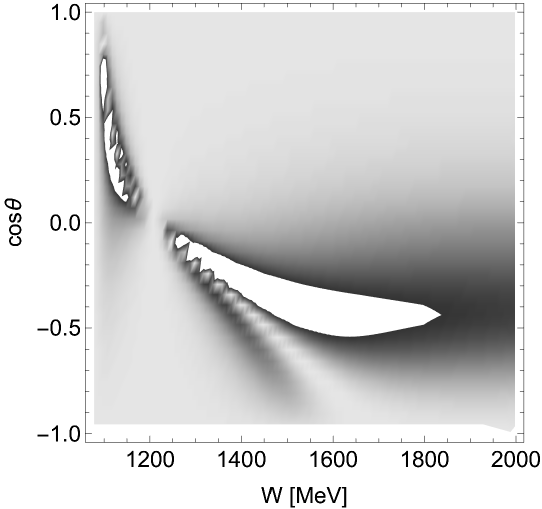}}{-0.4cm}{0.2cm}
\hspace{0.5cm}
\topinset{(c) $\Delta^{++}+\rho^0$}{\includegraphics[width=4.5cm]{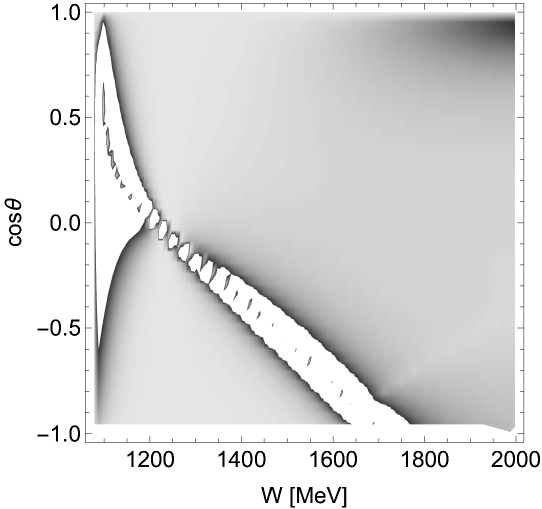}}{-0.4cm}{0.2cm}
\end{tabular}
\begin{tabular}{ccc}
\includegraphics[width=4.5cm]{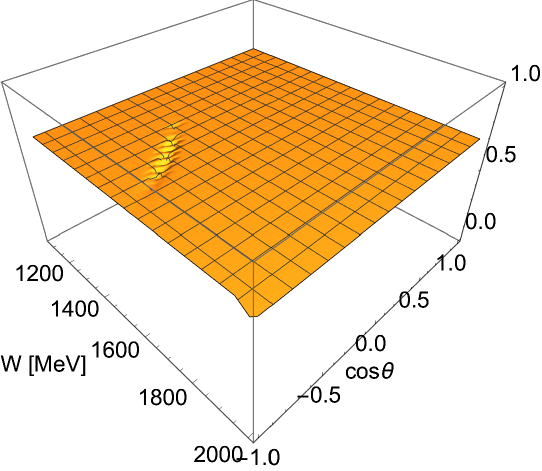}
\hspace{0.5cm}
\includegraphics[width=4.5cm]{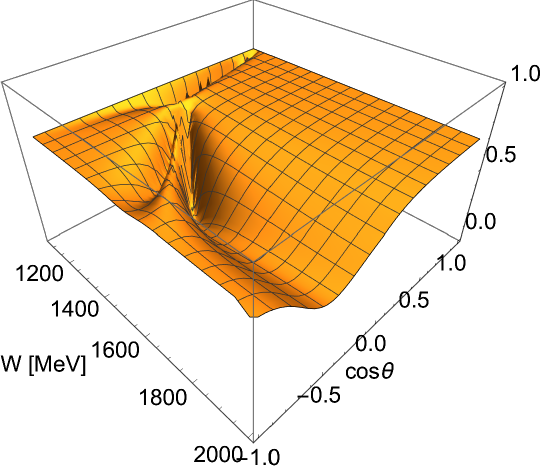}
\hspace{0.5cm}
\includegraphics[width=4.5cm]{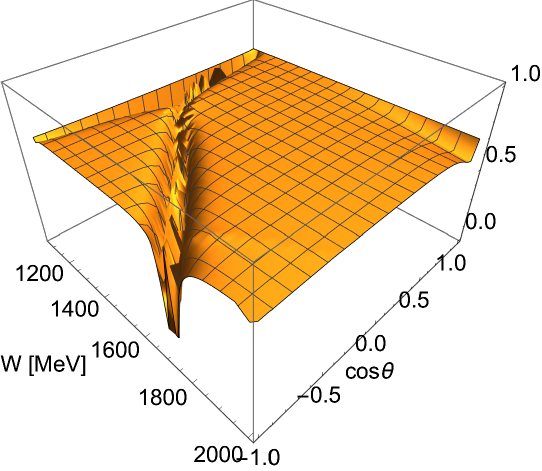}
\end{tabular}
\begin{tabular}{ccc}
\topinset{(d) $\Delta^0+n$}{\includegraphics[width=4.5cm]{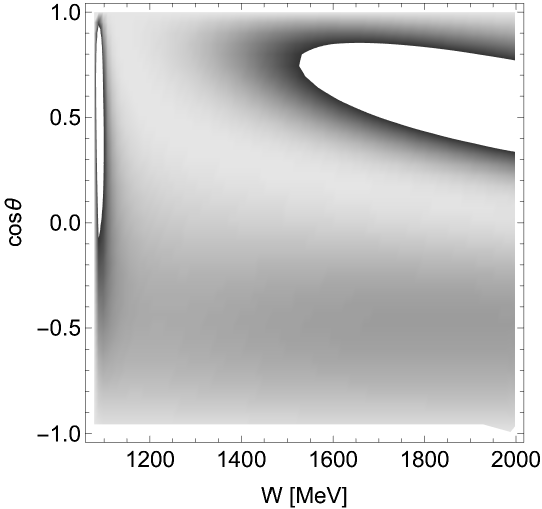}}{-0.4cm}{0.2cm}
\hspace{0.5cm}
\topinset{(e) $\Delta^0+\rho^0$}{\includegraphics[width=4.5cm]{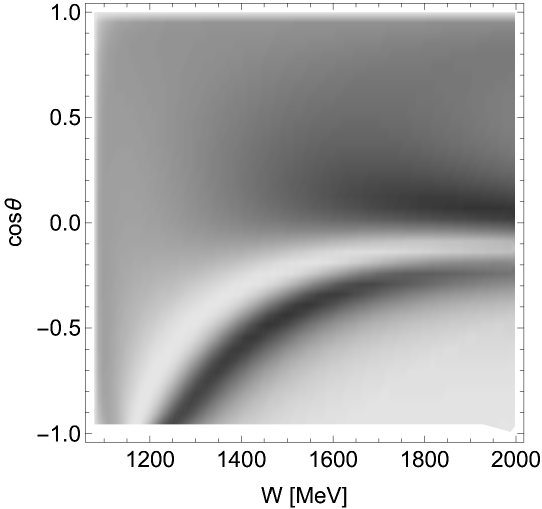}}{-0.4cm}{0.2cm}
\hspace{0.5cm}
\topinset{(f) $n+\rho^0$}{\includegraphics[width=4.5cm]{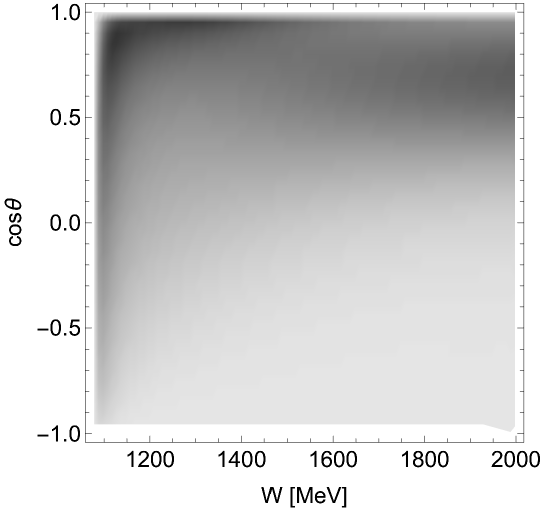}}{-0.4cm}{0.2cm}
\end{tabular}
\begin{tabular}{cccccc}
\includegraphics[width=4.5cm]{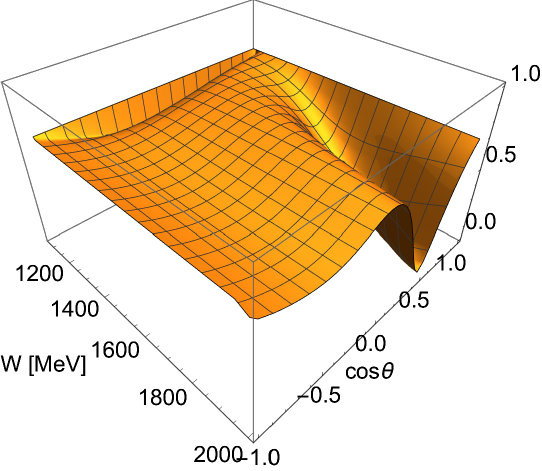}
\hspace{0.5cm}
\includegraphics[width=4.5cm]{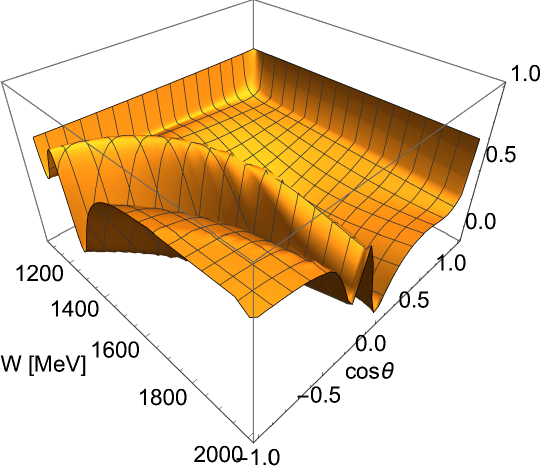}
\hspace{0.5cm}
\includegraphics[width=4.5cm]{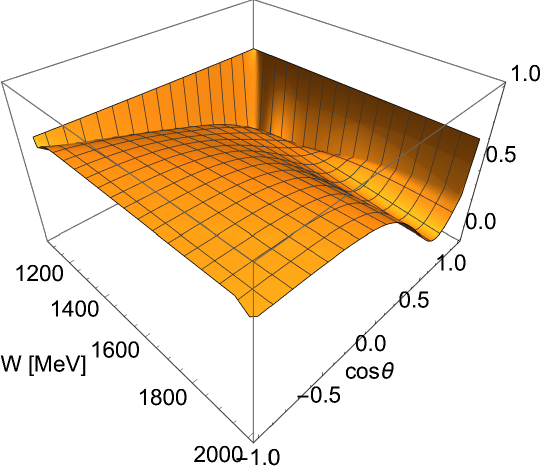}
\end{tabular}
\caption{\justifying (Color online) Similar to Fig.~\ref{FIG4}; $S_\mathrm{SDM}$ as a function of $W$ [MeV] and $\cos\theta$ for the (a) $\Delta^{++}+\Delta^{0}$, (b) $\Delta^{++}+n$, (c) $\Delta^{++}+\rho^0$, (d) $\Delta^0+n$, (e) $\Delta^0+\rho^0$, and (f) $n+\rho^0$ contributions.}
\label{FIG5}
\end{figure}
%FIGURE
To understand the specific roles of different channels in generating spin entanglement, we analyze entropy distributions for selected combinations of amplitudes in Fig.~\ref{FIG5}. For instance, when only $\Delta^{++}$ and $\Delta^0$ are included, the entropy remains almost flat across most kinematic regions, as depicted in Fig.~\ref{FIG5}(a), indicating strong entanglement between the same spin-$3/2$ states. We verified that the smaller-entropy region (spot-like) at $W\lesssim1.6$ GeV and $\cos\theta\lesssim0$ is mainly produced by the $\Delta^{++}$ contribution. However, the addition of neutron or $\rho^0$ exchange channels significantly alters the entropy structures, as shown in Fig.~\ref{FIG5}(b) and (c), making the entropy much smaller around that spot. This observation indicates that the $\Delta^{++}$ is disentangled from the different-spin background contributions, i.e., $n$ and $\rho^0$. Other combinations also exhibit nontrivial and distinctive entropy structures that differ from those with the $\Delta^{++}$ contribution, as shown in Fig.~\ref{FIG5}(d), (e), and (f). Thus, the von Neumann entropy effectively captures the transition from relatively pure to mixed spin states as background contributions become significant. Therefore, we can conclude that the combinations of different-spin particles tend to reduce entropy, leading to disentanglement and energy- and angular-dependent dynamics. In contrast, those of the same spin increase it due to stronger entanglement. Depending on the interference pattern, background amplitudes may either enhance mixed-spin entanglement or locally restore spin coherence, producing entropy dips.

Hence, our numerical analysis demonstrates that the von Neumann entropy derived from SDMs provides a novel and insightful observable for hadronic scattering processes, unlike traditional observables, such as cross-sections. The entropy probes internal spin correlations and coherence effects that would otherwise remain hidden.
%-------------------------------------------------
\section{Summary and future perspectives}
%-------------------------------------------------
In this work, we investigated the $\pi^+p$ elastic scattering process using an effective Lagrangian approach that incorporates $s$-, $t$-, and $u$-channel contributions involving $\Delta^{++}$, $\Delta^0$, neutron, and $\rho^0$. Tree-level scattering amplitudes were constructed and used to compute both total and differential cross-sections. The numerical results show qualitatively good agreement with existing experimental data, particularly around the $\Delta^{++}$ resonance peak near $W\approx1230$ MeV. Beyond conventional observables, we introduced von Neumann entropy as a quantum information measure to probe the spin dynamics of the scattering process. A 4$\times$4 SDM was built from the scattering amplitudes, based on canonical spin quantization in the c.m. frame, and partial traces were performed to extract 2$\times$2 reduced density matrices for the initial and final proton spin states. The corresponding von Neumann entropies quantify the degree of spin entanglement with other kinematic and internal variables.

The entropies exhibit a clear dependence on the c.m. energy $W$ and scattering angle $\theta$. In regions dominated by the $\Delta^{++}$ resonance, entropy is relatively low, leading to complex structures. In contrast, higher entropy values are observed when multiple channels interfere, particularly involving background contributions, such as neutron and $\rho^0$ exchange, indicating mixed spin states and enhanced quantum entanglement among different spin states. To further explore channel-specific effects, we decomposed the total amplitude into diagram subsets and computed entropy for each case. We observed that combinations involving only $\Delta$ resonances maintain almost flat entropy, whereas the inclusion of non-resonant backgrounds, such as $n$ and $\rho^0$, significantly alters the entropy structure, decreasing it. This again confirms the role of background amplitudes in generating spin decoherence, as evidenced by disentanglements. Overall, our results suggest that the von Neumann entropy extracted from SDMs serves as a novel and insightful observable in hadronic scattering. It complements traditional cross-section analysis by providing direct access to the quantum coherence and entanglement properties of spin subsystems. 

As observed in the valley-like dip structures, the von Neumann entropy of the spin system can serve as a quantum diagnostic tool for searching for resonances and exotic states in hadronic scattering. For instance, resonance formations dynamically favor specific spin states, thereby decreasing entropy and signaling a reduction in entanglement between the spin and other degrees of freedom, such as background contributions. Such entropy dips, as functions of $W$ and $\theta$, may reveal hidden resonant and exotic structures even when conventional observables are ambiguous. While challenging, experimental validation of the entropy structure is conceivable through future polarization-sensitive measurements. Observable quantities derived from the SDM, such as spin correlation coefficients, polarization transfer, or entanglement witnesses, could connect theory with data and make the framework operationally testable. Altogether, these enhancements would not only improve the quantitative precision of the model but also deepen its relevance to nonperturbative QCD and the emergent quantum features of hadronic systems.

Although SAID-based partial wave analyses provide data-driven amplitudes~\cite{Arndt:1990bp}, the present work focuses on a model-based construction of SDMs, enabling direct control over specific diagrammatic contributions and coherence structures. Nonetheless, we intend to complement the present formal development study with a numerical analysis using SAID partial-wave amplitudes. This will provide an explicit mapping between helicity amplitudes extracted from data and the canonical-basis amplitudes used here, serving as an external consistency check of our Wigner-rotation treatment and the two-independent-amplitude structure implied by parity. Such a comparison will be reported in a dedicated follow-up work. Related works will appear elsewhere.
%-------------------------------------------------
\section*{Acknowledgment}
%-------------------------------------------------
The author is grateful to the interesting discussions on von Neumann entropy and quantum entanglement with Tetsuo Hyodo (Tokyo Metropolitan University). This work is supported by grants from the National Research Foundation of Korea (NRF), funded by the Korean government (MSIT) (RS-2025-16065906). The partial support from the Institute for Radiation Science and Technology of PKNU (IRST/PKNU) is also acknowledged. 
%-------------------------------------------------
\section*{Appendix}
%-------------------------------------------------
In this Appendix, we provide technical details regarding the spin conventions, relation between canonical and helicity spinors via the Wigner rotation, implementation of parity constraints, and construction of reduced SDMs used in the text.
%-------------------------------------------------
\subsection{Canonical and helicity spinors}
%-------------------------------------------------
Canonical (or instant-form) proton spinors are defined by boosting rest-frame spin eigenstates to the desired momentum:
\begin{equation}
u_s(k) = S(\Lambda_k)\,u_s(0),
\qquad s=\uparrow,\downarrow ,
\label{eq:canonical_spinor}
\end{equation}
where $S(\Lambda_k)$ is the Dirac representation of the standard Lorentz boost
$\Lambda_k$. Helicity spinors are instead defined as eigenstates of 
$\mathbf{\Sigma}\cdot \hat{\mathbf{k}}$:
\begin{equation}
u_\lambda(k) = D^{1/2}[R(\hat{\mathbf{k}})]\,u_\lambda(0),
\qquad \lambda = \pm \tfrac12 ,
\label{eq:helicity_spinor}
\end{equation}
where $R(\hat{\mathbf{k}})$ is the rotation that aligns the $z$-axis with 
$\hat{\mathbf{k}}=\mathbf{k}/|\mathbf{k}|$.

The two spin bases are related by a momentum-dependent Wigner rotation $R_W$:
\begin{equation}
u_s(k) = \sum_{\lambda} 
D^{1/2}_{s\lambda}\!\left[R_W(k)\right]\,u_\lambda(k),
\label{eq:wigner_relation}
\end{equation}
where
\begin{equation}
R_W(k) = L^{-1}(k)\,R(\hat{\mathbf{k}})\,L(k),
\end{equation}
where $L(k)$ is the rotation-free boost to momentum $k$. In the c.m.\ frame, for scattering in the $xz$-plane, the Wigner matrix takes the explicit SU(2) form
\begin{equation}
D^{1/2}(R_W) =
\begin{pmatrix}
\cos \tfrac{\omega}{2} & -\sin \tfrac{\omega}{2} \, e^{-i\phi} \\
\sin \tfrac{\omega}{2} \, e^{i\phi} & \cos \tfrac{\omega}{2}
\end{pmatrix},
\end{equation}
where $\omega$ is the Wigner angle, and $\phi$ is the azimuthal scattering angle.
For collinear configurations ($\omega=0$), canonical and helicity bases coincide.
%-------------------------------------------------
\subsection{Parity constraints}
%-------------------------------------------------
In the helicity basis, parity invariance imposes the well-known relation
\begin{equation}
M_{\lambda_i \lambda_f}(\theta)
= \eta_P\, (-1)^{\lambda_i-\lambda_f}
M_{-\lambda_i, -\lambda_f}(\theta),
\label{eq:parity_helicity}
\end{equation}
reducing the number of independent $\pi N$ amplitudes to two. In the canonical basis, the scattering matrix elements read
\begin{equation}
M_{s_i s_f}
= \sum_{\lambda_i,\lambda_f}
D^{1/2}_{s_i \lambda_i}[R_W]\,
M_{\lambda_i \lambda_f}\,
D^{1/2\,*}_{s_f \lambda_f}[R_W],
\label{eq:canonical_transform}
\end{equation}
which generally do \textit{not} satisfy $M_{\uparrow\uparrow}=M_{\downarrow\downarrow}$
or $M_{\uparrow\downarrow}=-M_{\downarrow\uparrow}$ element-wise due to
$R_W(k)\neq1$ for noncollinear boosts.

Nevertheless, inserting Eq.~\eqref{eq:parity_helicity} into
Eq.~\eqref{eq:canonical_transform} shows that the rank of the canonical-basis amplitude matrix remains
\begin{equation}
\mathrm{rank}(\mathcal{M}) = 2,
\end{equation}
ensuring that only two physical invariant amplitudes exist, as required by
parity symmetry. Thus, the canonical and helicity descriptions are fully
equivalent at the level of physical degrees of freedom.
%-------------------------------------------------
\subsection{Reduced spin-density matrices and entropies}
%-------------------------------------------------
The $4\times4$ SDM for the proton spin system is
\begin{equation}
\rho_{s_i s_f; s'_i s'_f}
= \frac{1}{N} M_{s_i s_f} M^{*}_{s'_i s'_f},
\qquad
N = \sum_{s_i, s_f} |M_{s_i s_f}|^2 ,
\label{eq:rho_full}
\end{equation}
ensuring $\mathrm{Tr}[\rho_{SDM}]=1$. Reduced density matrices for initial and final spins follow from partial traces:
\begin{equation}
(\rho_i)_{s_i s'_i}
= \sum_{s_f} \rho_{s_i s_f; s'_i s_f},
\qquad
(\rho_f)_{s_f s'_f}
= \sum_{s_i} \rho_{s_i s_f; s_i s'_f}.
\label{eq:reduced_sdm}
\end{equation}
Their eigenvalues and the von Neumann entropy are
\begin{equation}
\lambda_{\pm}
= \frac12\left(1 \pm \sqrt{1-4\det\rho_{i,f}}\right),\,\,\,
S = -\sum_{\alpha=\pm} \lambda_\alpha \log \lambda_\alpha.
\end{equation}
For completeness, let
$\mathcal{H} = \mathcal{H}_A\otimes \mathcal{H}_B$. Then, we have
\begin{equation}
|\Psi\rangle = \sum_{i,j} c_{ij} |i\rangle_A |j\rangle_B,\,\,\,
\rho = |\Psi\rangle\langle\Psi|
= \sum_{i,j,k,l} c_{ij}c^*_{kl}
\left(|i\rangle_A\langle k|_A\right)
\otimes
\left(|j\rangle_B\langle l|_B\right).
\end{equation}

Tracing out subsystem $B$ yields
\begin{align}
\rho^A = \mathrm{Tr}_B(\rho)
= \sum_{m} \langle m |_B \rho | m \rangle_B \notag= \sum_{i,k}\left(\sum_{m} c_{im}c^*_{km}\right)
|i\rangle_A\langle k|_A ,
\end{align}
which is Eq.~\eqref{eq:reduced_sdm} in the present context. 

In the helicity basis, parity invariance implies that all spin information is encoded in two independent helicity amplitudes, say $H_1$ and $H_2$. Using the canonical transform,
\begin{equation}
M_{s_i s_f}
= \sum_{\lambda_i,\lambda_f}
D^{1/2}_{s_i \lambda_i}(R_W)\,
M_{\lambda_i \lambda_f}\,
D^{1/2 \,*}_{s_f \lambda_f}(R_W),
\end{equation}
and the helicity-parity constraint
\begin{equation}
M_{\lambda_i \lambda_f}
= \eta_P (-1)^{\lambda_i-\lambda_f}
M_{-\lambda_i,-\lambda_f},
\end{equation}
we can always express the canonical matrix as a linear combination
\begin{equation}
M = H_1\,\mathcal{U}_1 + H_2\,\mathcal{U}_2,
\end{equation}
where $\mathcal{U}_{1,2}$ are $2\times 2$ matrices built from Wigner rotation
elements. Hence $\mathrm{rank}(M)\le 2$. Since nontrivial $\pi N$ scattering
contains at least two independent amplitudes, we obtain
\[
\mathrm{rank}(M)=2,
\]
demonstrating that parity restricts the canonical amplitude to two 
independent components, consistent with the helicity formalism.

%-------------------------------------------------

%(^o^)
\end{document}